\documentclass[journal=aamick,manuscript=article]{achemso}

\usepackage{amssymb}
\usepackage{color}
\usepackage{graphicx}

\usepackage{siunitx}

\mciteErrorOnUnknownfalse

\title{Robust incorporation in multi-donor patches created using \phrase} 

\newcommand{\phrase}{atomic-precision advanced manufacturing}
\newcommand{\Phrase}{Atomic-precision advanced manufacturing}
\newcommand{\acro}{APAM}

\author{Quinn Campbell}
\affiliation{Center for Computing Research, Sandia National Laboratories, Albuquerque NM, 87185 USA}
\email{qcampbe@sandia.gov}
\author{Justine C. Koepke}
\affiliation{Sandia National Laboratories, Albuquerque NM, 87185 USA}
\author{Jeffrey A. Ivie}
\affiliation{Sandia National Laboratories, Albuquerque NM, 87185 USA}
\author{Andrew M. Mounce}
\affiliation{Sandia National Laboratories, Albuquerque NM, 87185 USA}
\author{Daniel R. Ward}
\affiliation{Sandia National Laboratories, Albuquerque NM, 87185 USA}
\altaffiliation{HRL Laboratories, LLC, Malibu, CA 90265, USA}
\author{Malcolm S. Carroll}
\altaffiliation{IBM Quantum, IBM T.J. Watson Research Center, Yorktown Heights, NY 10598, USA}
\affiliation{Sandia National Laboratories, Albuquerque NM, 87185 USA}
\author{Shashank Misra}
\affiliation{Sandia National Laboratories, Albuquerque NM, 87185 USA}
\author{Andrew D. Baczewski}
\affiliation{Center for Computing Research, Sandia National Laboratories, Albuquerque NM, 87185 USA}
\author{Ezra Bussmann}
\affiliation{Sandia National Laboratories, Albuquerque NM,  87185 USA}

\date{\today}

\begin{document}

\begin{tocentry}
\includegraphics[width=3.5in]{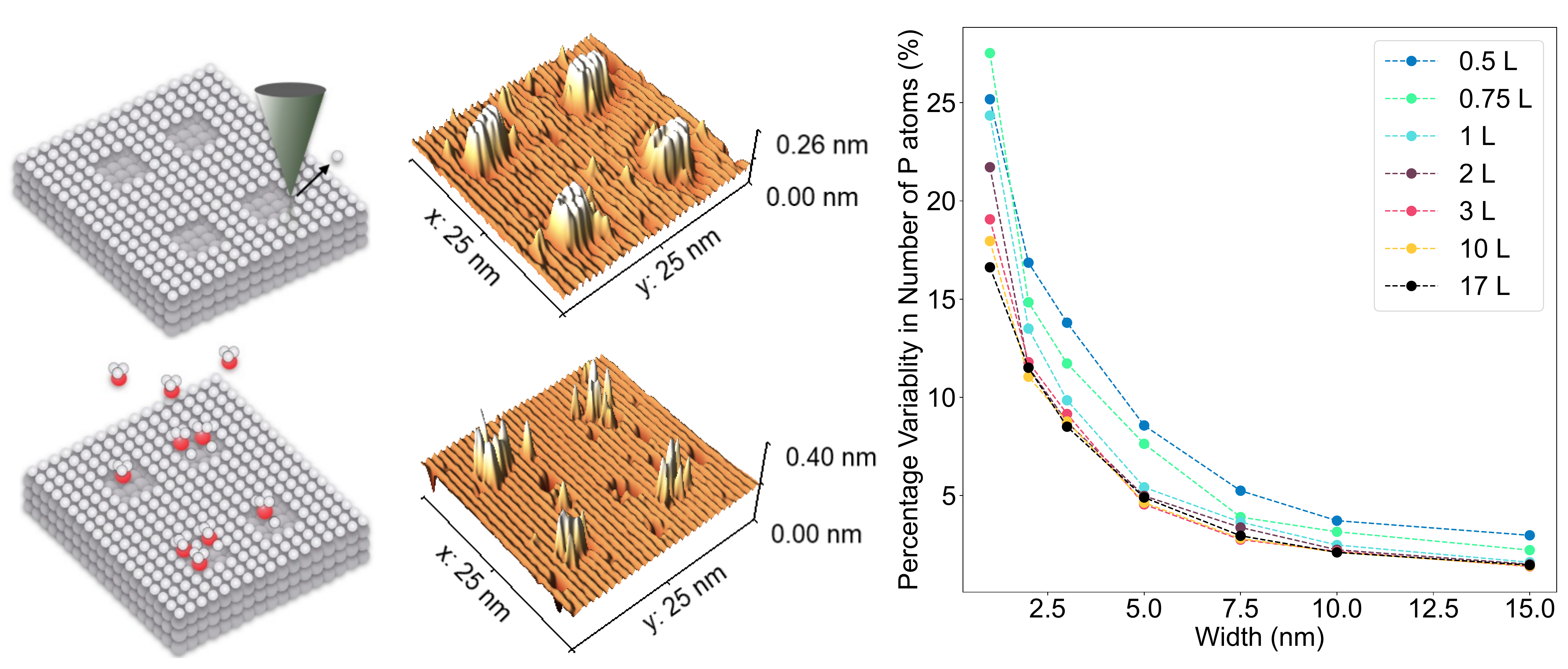}
\end{tocentry}

\begin{abstract}
\Phrase~enables the placement of dopant atoms within $\pm$1 lattice site in crystalline Si.
However, it has recently been shown that reaction kinetics can introduce uncertainty in whether a single donor will incorporate at all in a minimal 3-dimer lithographic window.
In this work, we explore the combined impact of lithographic variation and stochastic kinetics on P incorporation as the size of such a window is increased.
We augment a kinetic model for PH$_3$ dissociation leading to P incorporation on Si(100)-2$\times$1 to include barriers for reactions across distinct dimer rows. 
Using this model, we demonstrate that even for a window consisting of 2$\times$3 silicon dimers, the probability that at least one donor incorporates is nearly unity.
We also examine the impact of size of the lithographic window, finding that the incorporation fraction saturates to $\delta$-layer like coverage as the circumference-to-area ratio approaches zero.
We predict that this incorporation fraction depends strongly on the dosage of the precursor, and that the standard deviation of the number of incorporations scales as $\sim \sqrt{n}$, as would be expected for a series of largely independent incorporation events. 
Finally, we characterize an array of experimentally prepared multi-donor lithographic windows and use our kinetic model to study variability due to the observed lithographic roughness, predicting a negligible impact on incorporation statistics.
We find good agreement between our model and the inferred incorporation in these windows from scanning tunneling microscope measurements, indicating the robustness of \phrase~to errors in patterning for multi-donor patches.
\end{abstract}

\section{Introduction}
\label{sec:intro}

\Phrase~(\acro) is a suite of techniques for fabricating Si nanoelectronic devices for applications ranging from analog quantum simulators~\cite{georgescu2014quantum,prati2012anderson,prati2016band,le2017extended,dusko2018adequacy,altman2021quantum,wang21}, to qubits~\cite{buch2013spin,hill2015surface,pakkiam2018characterization,pakkiam2018single,he2019two,koch2019spin,bussmann2021apam}, to digital electronics~\cite{vskerevn2018cmos,ward2020atomic}.
\acro-fabricated devices are comprised of lithographically defined dopant structures with features that nominally have resolutions at the limit of a single-atomic site ($\pm 3.8$~\AA). 
While this resolution has been demonstrated for the placement of single dopant atoms~\cite{Schofield2003placement}, this manuscript examines the extent to which it persists for larger dopant structures in the presence of stochastic incorporation and lithographic imperfection.

\begin{figure}
\includegraphics[width=500pt]{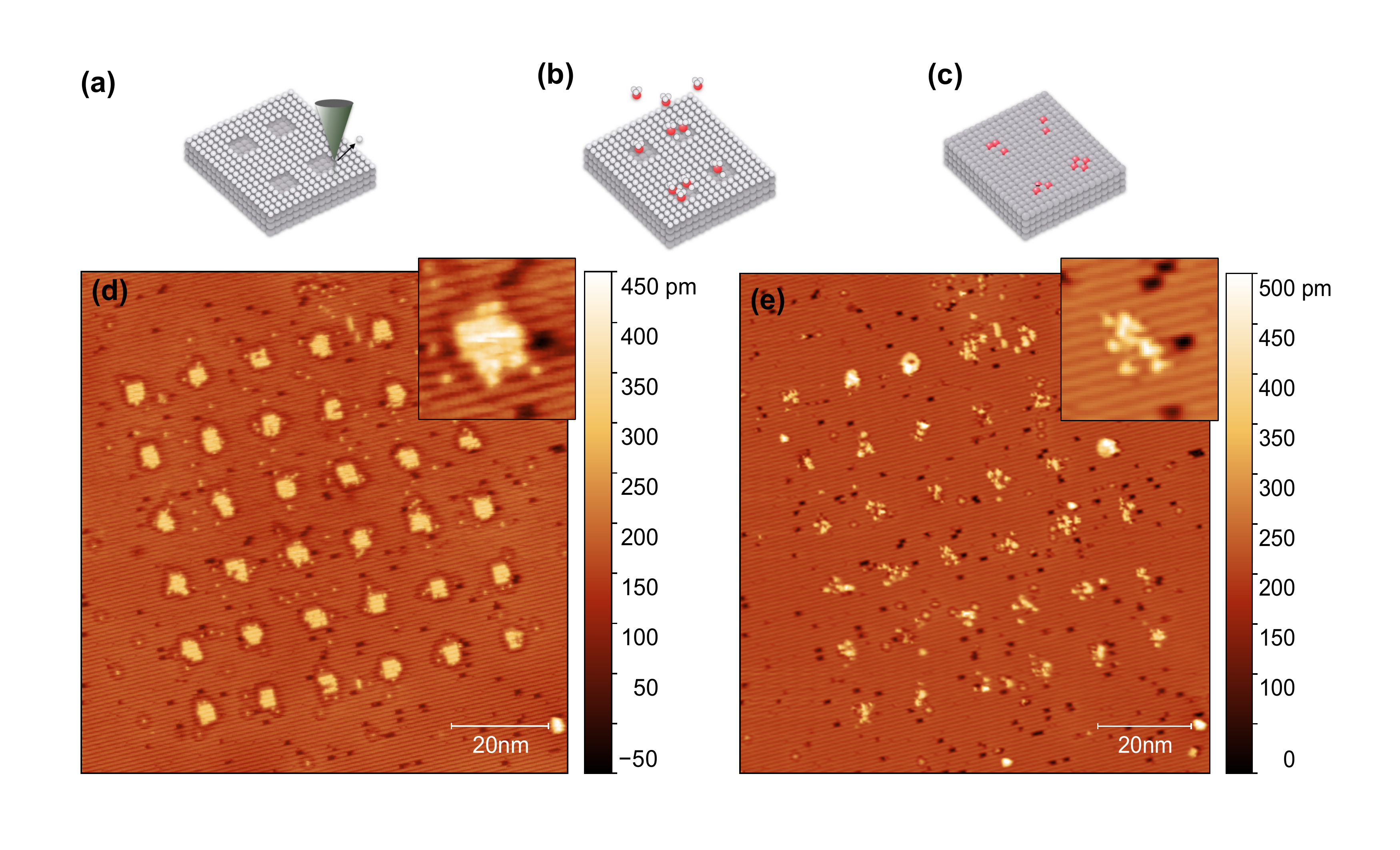}
\caption{The HDL-based APAM doping process steps are (a) patterning a lithographic template via atomic-precision HDL through a clean H-terminated H:Si(100)-2$\times$1 surface, (b) followed by exposure to PH$_3$ gas that selectively adsorbs onto Si in the HDL template, and finally (c) anneal-driven P incorporation within the template.
STM images throughout this process reveal variability in (d) the pattern around the edges of windows in the H resist, and (e) the outcome of the incorporation process, across an ensemble comprised of individual windows that are nominally patterned identically. HDL was performed using $+3.75$~V (sample), a $12.0$~nA STM tunneling current set point, and an electron dose of $6.0$ mC cm$^{-1}$. The STM images shown here were acquired with $-2.0$~V and $0.1$~nA set point. The bright features in panel (d) are depassivated Si regions (dangling bonds), while the bright features in panel (e) are thought to be primarily ejected Si displaced by P incorporation into the Si lattice. The insets in (d-e) are 10 nm squares. }
\label{fig:cartoon} 
\end{figure}

In contrast to doping via conventional ion implantation, \acro~methods introduce dopants through a window in an atomically thin lithographic resist (e.g., a passivating layer of hydrogen).
Fig.\ref{fig:cartoon}~(a) shows a window created by H depassivation lithography (HDL) in which a scanning tunneling microscope (STM) tip is used to remove the H-resist with atomic precision.
Then, as in Fig.\ref{fig:cartoon}~(b), dopant atoms are introduced by exposing the templated surface to a flux of a gas-phase precursor. 
During a subsequent anneal, dopants incorporate into the Si, Fig.\ref{fig:cartoon}~(c). Lateral dopant diffusion across the surface is bounded by the HDL template, while dopants are constrained to the surface layer by keeping the process temperature low ($<700$~K) to activate surface chemistry, while leaving bulk diffusion inactive.
STM images of the process reveal atomic-scale roughness in feature shapes at the HDL step, Fig.\ref{fig:cartoon}~(d) inset, and in the ultimate doping outcome, Fig.\ref{fig:cartoon}~(e).
The extent to which this is due to lithographic imperfection and/or stochastic incorporation kinetics is the question that we seek to answer.

It is worth emphasizing that this style of APAM fabrication relies on a relatively complicated sequence of surface chemical reactions for which the balance among the rates determine the ultimate stochastic incorporation outcome.
Each step requires activation over a kinetic barrier that is significant compared to energetic differences between the initial and final states.
Further, in many of the intermediate states, more than one reaction is simultaneously accessible enhancing the prospect for a random outcome. 
Thus we expect variability among samples in any ensemble depending on the probability of successful incorporation, $p$.
It has been empirically observed that delta layers achieve 1/4 P coverage and thus we can approximate a single incorporation event in a sufficiently large pattern as having a probability $p=0.25$.
To the extent that we can treat each possible P incorporation event as a single Bernoulli trial with probability $p$ (i.e., it either incorporates or does not with $p$ independent of other events) we expect the variance (standard deviation) in the number of dopants in a structure consisting of $n$ atoms to be $n p(1-p)$ ($\sqrt{np(1-p)}$).
For $p=1$ ($p=0$) the ensemble variability in $n$ will be negligible because the process will be deterministic, but for $p=0.5$ this variability will be strongest.
In this manuscript we predict that $p \approx 0.2$--$0.3$ for our process, and thus variability in the number of dopant atoms in a nominally $n$ atom structure is significant.

There are relatively few examples of studies of the ensemble variability of APAM-fabricated structures \cite{fuc10,wyr19}.
However, such studies are becoming increasingly relevant with the growing drive to scale up APAM-fabricated devices like quantum processors, analog quantum simulators, and classical electronic components/circuits. 
In each of these technologies there is significant interest in architectures based on structures comprised of few-dopant clusters for which variance in the number of dopants might substantially change the on-site energy of the dopant clusters. 
This could lead to substantial randomness in the properties of an array of such few-dopant clusters, making controlled quantum devices less achievable using this methodology.
Recent work in the field has selected the size of APAM engineered quantum dots specifically to avoid statistical variation in the number of dopants \cite{kiczynksi2022engineering}.
By better understanding incorporation kinetics we can treat variance in the number of incorporations in a small window as a new engineering degree of freedom.
This might enable robust device functionalities that continue to operate subject to this particular source of variability.

Here, we report the analysis of an ensemble of APAM windows and their incorporation statistics, relying on kinetic simulations of nanoscale lithographic features to understand the variability of doping outcomes for nominally identical nanostructures.
We predict that for small dimer windows, we should see at least one incorporation for all windows larger than 2$\times$3 silicon dimers in size.
For larger windows, we predict that the fraction of sites incorporated quickly saturates as the circumference-to-area ratio decreases.
Finally, we measure the experimental lithographic error on a number of 3 nm $\times$ 3 nm wide arrays.
We then use our kinetic model to predict the impact of this lithographic error on the system, finding good agreement with experimental results.
These indicate that lithographic errors tend to have a small impact on the final system compared to the inherent chemical reaction path dependent stochasticity.  
Our simulations shows that while varying incorporation pathways can lead to stochastic results for single donor arrays, at larger sizes, these APAM fabricated arrays are largely robust to both chemical and lithographic variation.
This implies that APAM multi-donor arrays can reliably be used in the construction of quantum dot devices in larger scale quantum computing and analog simulation. 

\section{Results}

To study the incorporation statistics of HDL-fabricated windows we augmented a Kinetic Monte Carlo (KMC) model first demonstrated by Ivie, Campbell, and Koepke \textit{et al.}\cite{ivie2021impact}. 
This prior work only considered reactions within a single dimer-row and thus the chemistry was expanded to include reactions that become possible in larger windows, particularly PH$_3$ dissociation reactions taking place across rows.
Warschkow \textit{et al.} demonstrated the first of these steps with a PH$_3$ molecule losing a hydrogen atom to a different silicon dimer-row (as opposed to losing the hydrogen to another dimer along the same row)~\cite{warschkow2016reaction}.
However, an exhaustive set of inter-row reactions has not been comprehensively explored.
We use density functional theory to calculate the reaction barriers for a large number of these reactions, with further details in Appendix A. 
We then include these reactions, along with the previously calculated reactions from Warschkow \textit{et al.}~\cite{warschkow2016reaction} in the KMC model from Ivie, Campbell, and Koepke \textit{et al.}\cite{ivie2021impact}.
More details of this kinetic model are given in Appendix B. 
The source code for these simulations is available on github \cite{kmccode}.  

\subsection{Incorporation in small multi-donor clusters}
\label{subsec:countable}
We first focus on small windows that only allow $\leq 5$ donors to incorporate.
For single-donor incorporation, it is common to create a window on one silicon dimer-row that is three dimers long~\cite{fuechsle2012single}.
This provides sufficient room for one PH$_3$ molecule to adsorb and then shed hydrogen to the surrounding silicon atoms before incorporating.
While this suggests a clear route to single-donor incorporation, at typical room-temperature dosing conditions there is a $>30\%$ probability that the associated P atom will fail to incorporate~\cite{Fuchsle2011thesis}, seemingly due to interference with another adsorbed PH$_3$~\cite{ivie2021impact}.
It is worth noting that other experiments have observed more favorable single-donor incorporation probabilities for smaller sample sizes~\cite{wyrick2021enhanced}.
These aren't inconsistent with data from larger samples, but they suggest that inference of the incorporation rate might be subject to systematic uncertainty that isn't yet fully understood.
This uncertainty notwithstanding, we should expect that expanding the three-dimer window used for single-donor incorporation will allow for more PH$_3$ atoms to adsorb and dissociate, making it much more likely that at least one donor will incorporate within the window.
We thus consider ``small'' windows as starting at three silicon dimers and expanding to include windows for which we expect $\leq 5$ donors to incorporate.

\begin{figure*}
 \includegraphics[width=\textwidth]{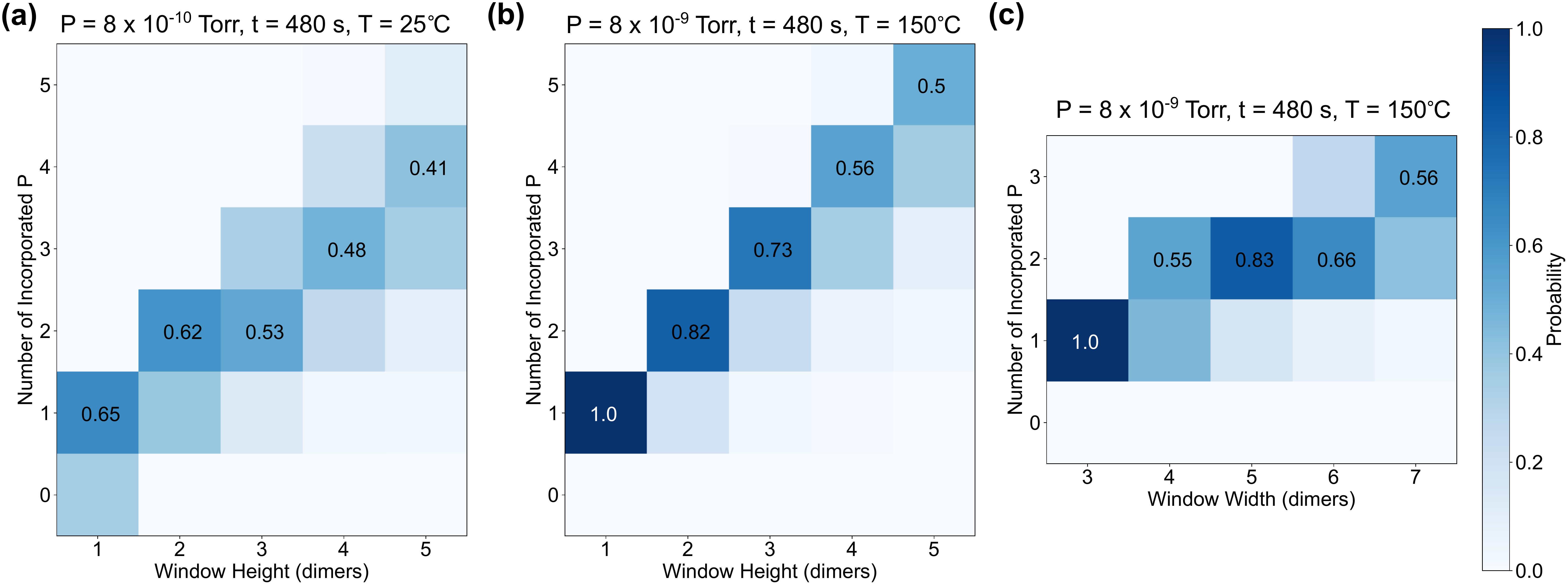}
\caption{Kinetic Monte Carlo predictions for the probability of incorporating different numbers of P atoms as a function of window size. In (a) and (b), we create a window three dimers wide and then extend it vertically across distinct dimer rows. Panel (a) shows the incorporation rate predicted for room-temperature dosing, while panel (b) shows the incorporation rate for a higher pressure and temperature that was previously predicted to lead to near-deterministic single-donor incorporation. In (c), we show predictions for a dimer window aligned along a single row with variable width for the same conditions as (b).}
\label{fig:countable_patch} 
\end{figure*}
	
In Fig.~\ref{fig:countable_patch}a, we show predictions from our kinetic model for the probability of incorporating 1 to 5 atoms in a window three dimers wide along a row, as a function of the number of silicon dimer rows included in the window (i.e., 1$\times$3, 2$\times$3, 3$\times$3, etc.).
The number of possible incorporations varies across different window sizes. 
For both a 2$\times$3 and 3$\times$3 window, the most likely number of P incorporations is two, with a probability of 62\% and 53\%, respectively. 
For all windows larger than 2$\times$3, the range of possible incorporations is at least three.
Notably, even at the room-temperature dosing conditions, we predict only a neglible 0.2\% chance of zero P atoms incorporating in any windows more than 2 dimers rows and 3 dimers within a row.
This confirms what one might intuitively expect-- that creating larger windows is an effective strategy for guaranteeing that every element in an array comprised of multi-donor patches has at least one donor.
Our predictions suggest a lower bound on the window dimension for which this strategy will work.

Based on conditions predicted to lead to deterministic incorporation in our previous work, we also explore the same window sizes while using a higher temperature of \SI{150}{\celsius} during dosing.
We predicted this arrangement would lead to deterministic incorporation for a single donor in the case of a 1$\times$3 window~\cite{ivie2021impact}.
As shown in Fig.~\ref{fig:countable_patch}b, however, this elevated temperature dosing is not predicted to remain deterministic at larger patch sizes.
While the number of expected incorporation events reliably shifts up, the probability of achieving as many donor incorporations as the number of depassivated sites would suggest decreases as the size of the window increases.
This stochasticity can be entirely attributed to the new reactions across dimer rows introduced in Sec.~\ref{sec:multi-row-rxns}.
Without these reactions, a 5$\times$3 dimer window would be exactly equivalent to five separate 1$\times$3 windows and five deterministic incorporation events would be expected at the elevated temperature dosing conditions under consideration.
This shows that reactions across dimer rows are competitive with reactions along dimer rows and further that these reactions can lead to site blocking, preventing the dissociation of some of the adsorbed PH$_3$ molecules. 

We finally examine windows oriented along a single dimer-row, but with different widths in Fig.~\ref{fig:countable_patch}c.
We again use the elevated temperature conditions associated with the deterministic incorporation of a single donor for a 1$\times$3 window. 
We again see, however, that the deterministic nature of incorporation is limited to the 1$\times$3 window providing exactly enough space for one molecule to dissociate.
As the width of the window increases, more than one PH$_3$ has room to dissociate within it.
This might lead to multiple incorporations, but interference among adsorbed molecules might also prevent all incorporations but one.
Notably, a 1$\times$6 window does not behave the same as two 1$\times$3 windows, there is still a 7.6\% chance that only a single donor will incorporate.
This stochasticity brings both the promise of more potential incorporations and the danger of fewer. 
A 1$\times$6 window has a 26.8\% chance of three donors incorporating if the adsorption and dissociation proceeds just right, which would not be possible in two separated 1$\times$3 windows.
Critically, this disorder does not lead to any configurations where zero incorporation events are predicted. 

\subsection{Size-dependence of incorporation in larger multi-donor clusters}
\label{subsec:size-dep}
\begin{figure*}
 \includegraphics[width=\columnwidth]{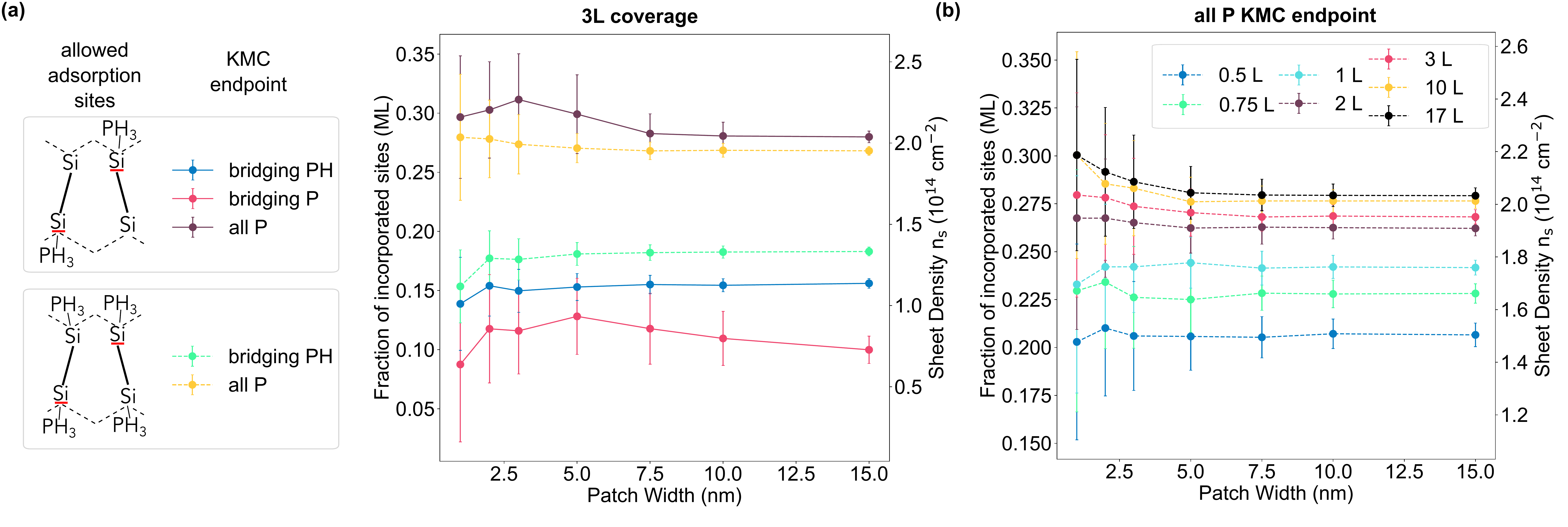}
\caption{(a) The fraction of the surface that is incorporated P (based on the choice of proxy within our kinetic model) as a function of the square patch width using a constant 3 L dose. 
(b) The fraction of the surface incorporated as a function of different langmuir doses when using an all P proxy for incorporation and phosphine allowed to adsorb on all silicon sites.
Error bars are extracted from standard deviations of the number of incorporations. 
}
\label{fig:width_size} 
\end{figure*}
One important question facing kinetic models of the APAM doping process is how to assign surface dissociation outcomes as leading to P incorporation.
While the initial steps of PH$_3$ dissociation on the silicon surface have been well mapped out by Warschkow \textit{et al.}~\cite{warschkow2016reaction} and extended in Appendix A, modeling the full substitutional incorporation process once the P atom moves below the surface is computationally prohibitive.
In our previous work, we used a bridging PH fragment as a proxy for incorporation. 
This was because its significant adsorption energy makes negligible the likelihood of the P atom desorbing during subsequent anneals~\cite{ivie2021impact}. 
We further showed that this choice of incorporation proxy led to good agreement with the experimentally inferred incorporation rate based on STM images taken after an incorporation anneal.
An analogous choice of bridging BH and BCl as incorporation proxies has also aligned with electrically measured incorporation rates for diborane and boron trichoride precursors~\cite{campbell2021model,campbell2022reaction}.

Nevertheless, this choice of incorporation proxy raises an important question-- where does the final H on the bridging PH molecule go before incorporation?
If it is shed onto the surface it could prevent the further dissociation of nearby PH$_3$ molecules and fragments thereof, potentially lowering the overall incorporation rate.
On the other hand, if the H is abstracted toward the surface within the post-anneal growth process and the P subsequently incorporates, then this process could likely also be extended to a less stringent proxy of PH$_2$. 
To investigate this question, we examine the size dependence of APAM incorporation rates using multiple different proxies for incorporation. 

Fig.~\ref{fig:width_size}a compares the incorporation rates predicted using different proxies for incorporation and allowed adsorption sites as a function of the size of a square depassivated window with dosing conditions providing a 3 Langmuir (L) dose. 
We predict that once the lithographic window for atomic-precision exceeds $\approx$ 5 nm, the mean incorporation rate per unit area is practically constant for most incorporation proxies.
The higher degree of change in the mean incorporation fraction at patch widths below 5 nm can likely be attributed to edge effects being relatively more important in smaller systems, where the circumference-to-area ratio is higher. 
As expected, the asymptotic density of incorporations is lower when using a bridging P atom as a proxy for incorporation instead of a bridging PH molecule. 

We note that the fraction of a monolayer covered by incorporations for both the bridging PH and bridging P proxies (0.1 ML and 0.15 ML, respectively) is below the 0.25 ML figure from the literature.
To investigate the limits of varying the incorporation proxy, we also consider a proxy for incorporation in which all of the P on the surface eventually incorporates, i.e. even P atoms that are locked in PH$_3$ or PH$_2$ configurations will be counted as incorporating.
This is an overestimate of the actual number of incorporations as we are assuming perfect incorporation of molecules that have not completely dissociated and are likely to migrate. 
In this case, we predict much higher incorporation rates with an asymptotic value of $\approx$ 0.28 ML.

We next examine an additional assumption that has been used throughout our kinetic model. 
Silicon surface dimers tilt up and down, and Ivie, Campbell, and Koepke, \textit{et al.} have shown using DFT that the lower side of the dimer is more favorable for PH$_3$ adsorption~\cite{ivie2021impact}.
Our model has therefore assumed that PH$_3$ molecules only adsorb on this lower side, an approach which led to excellent agreement with experiment for single-donor APAM processes~\cite{ivie2021impact}.
The DFT calculations, however, are clear that PH$_3$ favorably adsorbs on both ends of the dimer even if the lower side is $\approx$ 0.2 eV more favorable. 
Additionally, in practice, the dimer tilt flips rapidly on the silicon surface, and there is no clear way of knowing \textit{a priori} which tilt will be available when a PH$_3$ molecule adsorbs on the surface.
Thus we examine a KMC run where PH$_3$ is allowed to adsorb with equal probability on both sides of the dimer, using both bridging PH and all P atoms as a proxy for incorporation.
We predict that this increases the total number of incorporations when a bridging PH proxy is used, but slightly decreases the incorporation rate when counting all P on the surface.
We attribute this decrease of total P to making it slightly easier to form bridging PH features, whose dissociation of hydrogen to nearby dimers can block further adsorption, lowering the number of PH$_2$ fragments on the surface. 
For the bridging PH case, we predict an asymptotic coverage of $\approx$ 0.18 ML, and 0.27 ML coverage for all P on the surface.

The amount of phosphorus on the surface is highly dependent on the initial dose. 
In Fig.~\ref{fig:width_size}b, we show how the surface coverage changes as a function of the dose when using an all P incorporation proxy and allowing phosphine to adsorb on all lattice sites. 
For all of these systems, we hold the dosing time constant at 540 seconds and change the dosing pressure to achieve the stated Langmuir dose. 
Matching intuition, the amount of the phosphorus on the surface increases along with the dose, eventually saturating at around 10 L. 

For comparison, several prior experiments estimated the P incorporation rates by closely examining P adsorption and incorporation chemistry on clean Si$(100)$, finding a temperature and pressure-dependent saturation P density $0.37$ML (T$=300$K) that drops to $0.25$ML (T$=700$K).~\cite{yu86,lin99,scho06} Uncertainties are roughly $0.05$ML owing to the challenge of accurately calibrating surface analytical tools, which included, e.g. X-ray photoelectron spectroscopy (XPS) and Auger electron spectroscopy (AES), and STM. 

That incorporation rates are sensitive to T and PH$_3$ pressure (dose rate) reemphasizes a role that kinetics plays in the doping process. For $T=300$K and various PH$_3$ pressures $<10^{-7}$ Torr, the surface saturates at a P density of $0.37\pm0.05$ML, according to Yu and Lin’s XPS and Auger electron spectroscopy studies.~\cite{yu86,lin99} This coverage is consistent with STM data reported indepedently by e.g. Wang, Oberbeck, Shen, and Schofield that indicates a short-range ordered c$(4\times2)$ phase coexisting with a p$(2\times2)$ phase, as first recognized by Schofield.~\cite{scho06,wang94a,shen02,obe02} Considering these two forms of order, we expect a coverage between the $0.25$ML of the c$(4\times2)$ phase, and the $0.5$ML of p$(2\times2)$.~\cite{shen02, obe02,scho06} With increasing T, the trend reported by Yu and Lin, also reflected in subsequent electronic transport experiments from McKibbin, is a monotonic decrease such that by $T=700$K, the saturation density drops to $0.25\pm0.05$ML.~\cite{mck09,mck14}

\begin{figure}
 \includegraphics[width=0.5\textwidth]{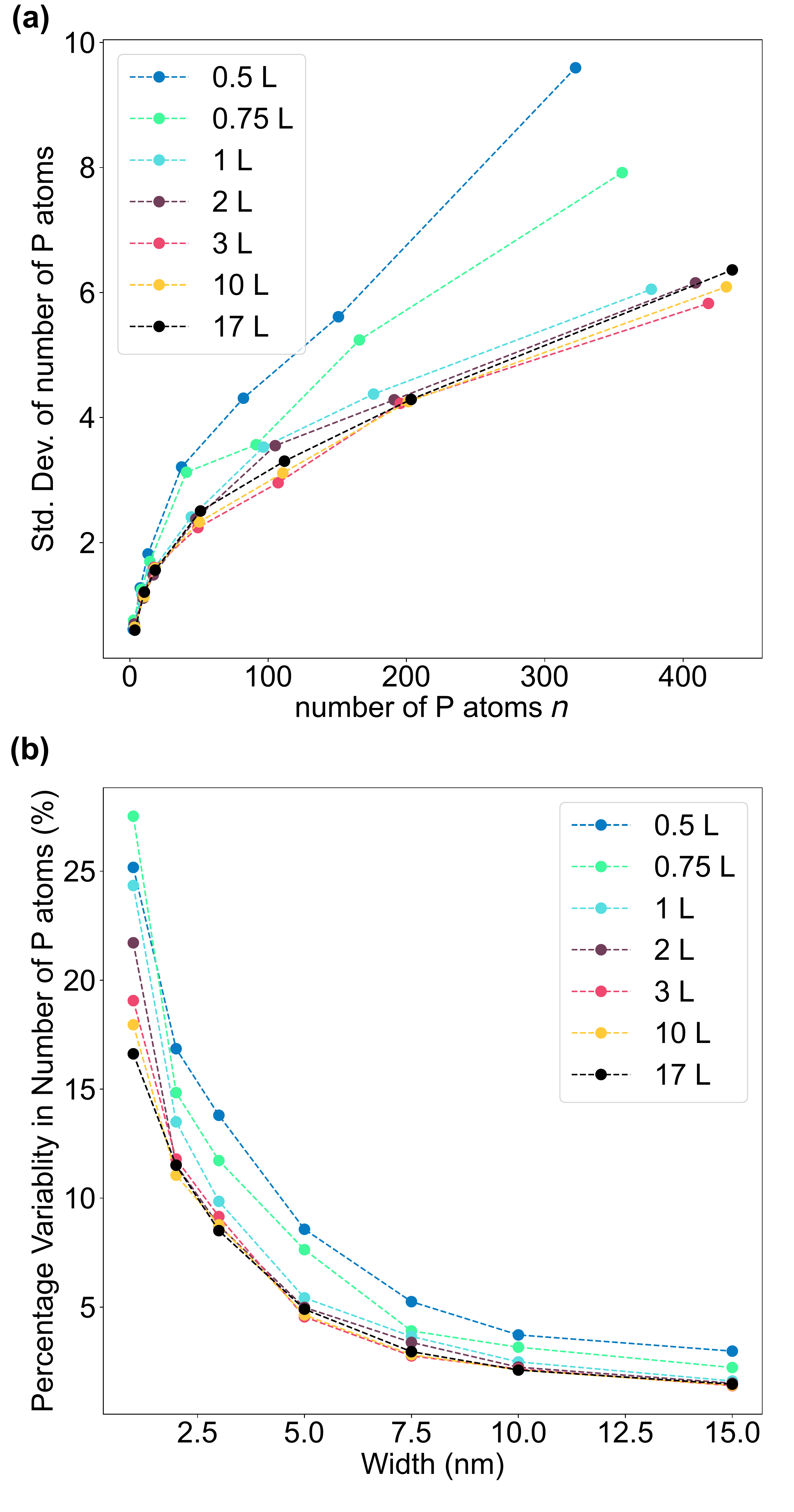}
\caption{(a) The standard deviation of the number of phosphorus atoms $n$ in a given APAM patch. 
We note a $\sqrt{n}$ scaling, supporting the intuition that these systems can be treated, to a first approximation, as a series of independent Bernoulli trials for each incorporation event.
(b) The variability in the total number of phosphorus atoms on the surface as a function of the width of the APAM patch.
As the size of the patch increases, the variability significantly decreases. 
}
\label{fig:size_std_dev} 
\end{figure}

Finally, in Fig.~\ref{fig:size_std_dev}a, we predict the relationship between the number of phosphorus atoms $n$ in a given APAM array, and the measured standard deviation in this number.
We find a square root relationship in the error, supporting the intuition developed in Sec.~\ref{sec:intro} that multi-donor arrays can be qualitatively treated as a series of independent Bernoulli trials for a fixed probability of incorporation $p$ and a standard deviation $\sqrt{np(1-p)}$.
Since all of these simulations were done assuming perfect lithography, this relationship can be solely attributed to the stochastic nature of the phosphine decomposition pathway. 
The slope of the standard deviation curves varies as a function of the dose, which can then be related back to the probability of success in a Bernoulli trial via the formula for standard deviation.
As the dose increases, the slope of the standard deviation curve stabilizes, indicating that the probability of incorporation saturates for doses greater than 1 L, matching intuition. 
These results indicate that the stochastic chemistry based error for a given APAM array will be a significant fraction of the total incorporated atoms at low system sizes such as single donor features, as shown in Fig.~\ref{fig:size_std_dev}b. 
As the size of the APAM array increases, however, the error will saturate and become a much smaller fraction of total incorporations, making these large systems highly robust to the stochastic nature of phosphorus incorporation.

\subsection{Impact of Lithographic Roughness}
\begin{figure*}
 \includegraphics[width=\textwidth]{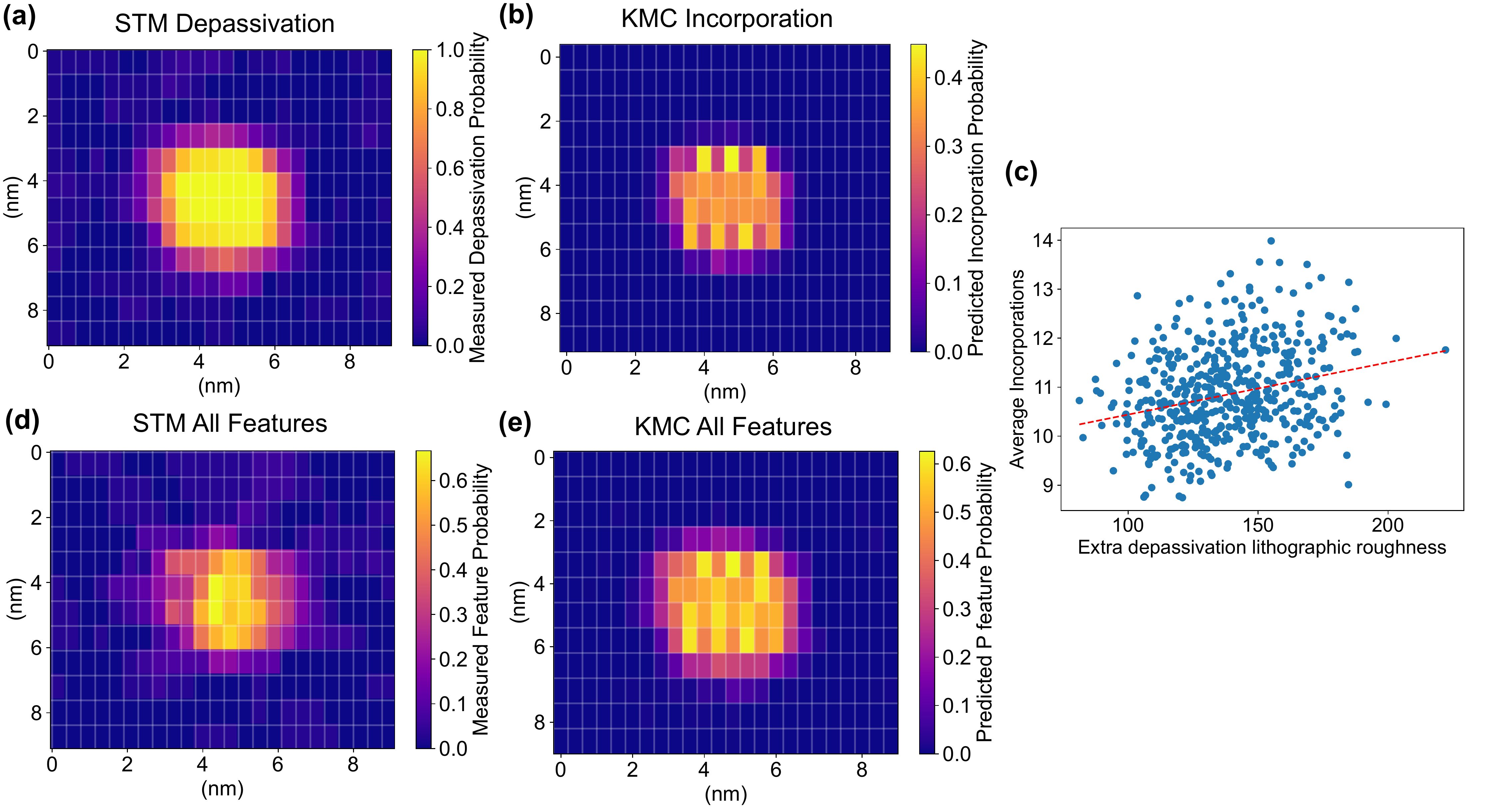}
\caption{(a) The experimentally measured likelihood of a site being depassivated after STM lithography. Values are averaged from 25 samples. 
(b) The predicted location of incorporated P in our kinetic model using the depassivation probabilities of panel a to generate sample inputs. The disorder is noticeably less than the total lithography error.
(c) No strong correlation is seen between the lithographic roughness of the depassivated window (as defined in Eq.~\ref{eq:roughness} and the average number of incorporations observed for the window.
(d) The experimentally measured likelihood of finding a feature after dosing the sample with PH$_3$ and annealing.
(e) The predicted probability of finding any P atom on the surface at a given site. The results suggest that the amount of P on the surface within our kinetic model matches experimental surface features well. }
\label{fig:litho_roughness} 
\end{figure*}
In a real world experiment, quite often the STM lithography will not be perfect, yielding windows with too many or (less often) too few sites depassivated. For example, note the varying shape and size of depassivated windows across the array in Fig.~\ref{fig:cartoon}d. 
In Fig.~\ref{fig:litho_roughness}a, we have averaged the actual depassivation rate for each dimer across 36 attempts at a 3 $\times$ 3 nm window using STM lithography. We calculate the image in Fig.~\ref{fig:litho_roughness}a using thresholding to identify depassivated dimers, yielding a binary image 
followed by calculating mean of the set of 36 windows. Since depassivated dimers protrude from the surrounding passivated surface by a peak height of roughly 150 pm, we identify a dimer as depassivated if it's median height exceeds a threshold value, chosen as 100 pm relative to the surrounding passivated surface. The fidelity of thresholding is comparable to more labor intensive visual-manual feature identification. In Fig.~\ref{fig:litho_roughness}a, the soft transition from 0 to 1 over a range of about 1 nm in any direction indicates lithographic roughness that is typical in HDL.~\cite{wyr19,wang21,kiczynksi2022engineering} While techniques such as feedback lithography could be used to improve the sharpness of the defined lithographic window ~\cite{hersam2000silicon,wyrick2018atom}, this represents a reasonable STM lithography run that is similar to recent works utilized to fabricate quantum electronics.~\cite{wyr19,wang21,kiczynksi2022engineering}

Using our kinetic model, we predict that these errors in depassivation are often self-correcting, i.e. do not lead to significant deviations in the total number of incorporations for a given window.
We investigate this by creating 500 depassivation patterns for our kinetic model based on the the empirical probability map in Fig.~\ref{fig:litho_roughness}a.
This approach assumes that the probability of depassivation at each dimer is independent, when in reality they are likely at least somewhat correlated. 
This model nevertheless serves as a good first approximation to real world systems, as any given configuration contains an average of up to 20 lattice sites depassivated that were not originally intended. 
In Fig.~\ref{fig:litho_roughness}b, we demonstrate the predicted end location of incorporations, showing a sharp drop off in incorporations beyond the original window. 
While typical lithographic errors may be non-trivial beyond the intended window, these errors are largely self-correcting: the errors do not give the PH$_3$ enough room to fully dissociate and thus are less likely to incorporate. 

We further demonstrate the robustness of the incorporation to lithographic error by investigating the relationship between the ``lithographic roughness'' of a system and the number of incorporations.
We define the total lithographic roughness $\sigma$ for a given configuration as 
\begin{equation}
    \sigma = \sum_i \sqrt{(x_i^{\rm e}-x^{\rm w}_{i})^2 + (y_i^{\rm e}-y^{\rm w}_{i})^2},
    \label{eq:roughness}
\end{equation}
where $i$ is a counter over the set of depassivation errors, $x^{\rm e}$ and $y^{\rm e}$ are the $x$ and $y$ locations of the error respectively, and $x^{\rm w}$ and $y^{\rm w}$ are the $x$ and $y$ locations of the nearest lattice site within the intended lithographic window. 
This definition of roughness means that in systems where depassivation errors occur close to the originally intended window, the score would be lower than in systems where the errors are far away from the intended window.
As shown in Fig.~\ref{fig:litho_roughness}c, we find a low correlation between the total lithographic roughness $\sigma$ and the average number of incorporations for a system with an r$^2$ value of just 0.06. 
While there is a slight positive correlation of incorporations with increased lithographic roughness, this variation is almost entirely swamped by the typical kinetic variation in total numbers of incorporations due to the uncertain chemical pathway for PH$_3$ dissociation. 
We thus conclude that these atomic-precision processes in these systems are robust to typical lithographic error.

Finally, we demonstrate agreement between experiment and prediction in the location of post-doping incorporation-related features on the surface. Using STM data in Fig.~\ref{fig:cartoon}e, acquired after dosing and annealing the lithographic windows used to generate the depassivation probability map in Fig.~\ref{fig:litho_roughness}a, we threshold and count all features with height greater than 70 pm above the surrounding hydrogen passivation, then calculate a mean heatmap of P incorporation-related features in Fig.~\ref{fig:litho_roughness}d. 
These features likely comprise a wide variety of features, including varying Si:P features and ejected Si adatoms on the surface. 
We identify all of these as likely to be P based features, not making strong distinctions between different PH$_x$ fragments, on the assumption that ejected Si adatoms represent P incorporating below this surface.
This lack of distinction is both based on the limited resolution of the STM images and the desirability of developing a general sense of where P atoms are located on the surface, regardless of eventual incorporation. 
This is somewhat speculative; some amount of silicon adatoms will migrate both in and out of the window, and perhaps a few residual dangling bonds will be captured.
Nevertheless, this provides a reasonable first-order estimate of where P features are located within a window. 
We then use our kinetic model to predict the location of all P atoms on the surface given an identical dosing and annealing process, as shown in Fig.~\ref{fig:litho_roughness}e. 
The prediction matches the experimentally measured location well, with both showing peak probabilities in the center of the depassivated window around 60\%.
They both also show a steep dropoff in P features around the edge of the lithographic window, with only locations 1-2 lattice sites away from the intended window seeing significant levels of P buildup. 
In contrast to the predicted incorporations in Fig.~\ref{fig:litho_roughness}b, however, the edge is less sharp, demonstrating that the need for PH$_3$ dissociation is the main physical mechanism for atomic-precision in response to lithographic error.
This provides a useful check on our kinetic model, giving confidence to our earlier predictions of the robustness of multi-donor incorporation. 
	
\section{Summary}	

Using kinetic simulations of the APAM doping process facilitated by hydrogen desorption lithography, we have predicted that multi-donor arrays are robust to both chemical and lithographic variation. 
Our results suggest that for small lithographic windows, even a window three silicon dimers wide and two dimer rows tall is sufficient to guarantee at least a single P incorporation.
They also suggest that at window sizes larger than 5 nm the incorporation per unit area remains relatively constant, defining the scale at which edge effects become less pronounced.
We further predict that the behavior of the system can, to a first order, be taken as a series of independent Bernoulli trials for each incorporation event.
We examine typical lithographic variation in creating APAM windows, and simulations suggest that little correlation exists between the ``roughness'' of lithography and the resulting incorporations.
This suggests that lithographic errors are largely self-correcting.
Finally, we corroborate our model by comparing the predicted location of all PH$_x$ fragments on the surface to STM images.

Our results suggest that APAM multi-donor arrays are robust to typical sources of error in both lithographic and chemical processes.
Thus APAM fabrication should be reliable for scalable device applications where more than a single donor can be used in individual elements (e.g., arrays of quantum dots). 

\begin{acknowledgement}
We gratefully acknowledge useful conversations with 
Scott Schmucker,
Rick Muller, 
Peter Schultz,
and Joe Simonson.
This work was partially supported by the Laboratory Directed Research and Development program at Sandia National Laboratories under project 213017 (FAIR DEAL) and project 226354. 
This work was also performed, in part, at the Center for Integrated Nanotechnologies, a U.S. DOE, Office of Basic Energy Sciences user facility.
Sandia National Laboratories is a multi-mission laboratory managed and operated by National Technology and Engineering Solutions of Sandia, LLC, a wholly owned subsidiary of Honeywell International, Inc., for DOE’s National Nuclear Security Administration under contract DE-NA0003525.
This paper describes objective technical results and analysis.  
Any subjective views or opinions that might be expressed in the paper do not necessarily represent the views of the U.S. Department of Energy or the United States Government.
\end{acknowledgement}

\begin{suppinfo}
Appendices are included in the Supporting Information concerning (A) DFT calculations across multiple silicon dimer rows, (B) Kinetic Monte Carlo details, and (C) 3$\times$3 nm heatmaps of incorporation.
\end{suppinfo}

\bibliography{new_refs}

\clearpage

\setcounter{section}{0}
\setcounter{page}{1}
\makeatletter

\section*{Appendix A: DFT calculations across multiple silicon dimer rows}
\setcounter{equation}{0}
\setcounter{figure}{0}
\renewcommand{\theequation}{A\arabic{equation}}
\renewcommand{\thefigure}{A\arabic{figure}}
\label{sec:multi-row-rxns}
\section{Electronic Structure Calculation Details}
All total energy calculations are performed using the plane wave {\sc quantum-espresso} software package.\cite{giannozzi2009quantum} 
To compute reaction barriers between configurations we use the Nudged Elastic Band (NEB) method, also implemented in {\sc quantum-espresso}. 
We use norm-conserving pseudopotentials from the PseudoDojo repository~\cite{van2018pseudodojo} and the Perdew-Burke-Ernzerhof exchange correlation functional.~\cite{perdew1996generalized}
We use kinetic energy cutoffs of 50 Ry and 200 Ry for the plane wave basis sets used to describe the Kohn-Sham orbitals and charge density, respectively.
We use a 2$\times$2$\times$1 Monkhorst-Pack grid to sample the Brillioun zone.\cite{monkhorst1976special}

We perform all adsorption energy calculations on the 4$\times$4 supercell of a seven-layer thick Si(100)-2$\times$1 slab with a 20 \AA~vacuum region, where a single unit cell has a size of 3.87 \AA$\times$3.87 \AA. 
Matching the experimentally measured silicon structure, we model the silicon surface as antisymmetric with alternating buckled silicon dimers. 
On the other end of the slab, the dangling bonds of the silicon are passivated with selenium atoms to prevent spurious surface effects. 
Selenium was determined to be optimal for achieving this purpose with minimal strain.
The bottom four layers of the slab are frozen and the geometry of the surface layers along with the adsorbate are relaxed until the interatomic forces are lower than 50 meV/\AA.

\section{Reaction barriers of reactions across multiple silicon dimers}
We use Density Functional Theory (DFT) calculations to predict the reaction barriers of various reactions taking place across dimers. 
Given the distance between silicon dimer rows in the (100)-2$\times$1 reconstruction is only $\sim$ 3.5 \AA, reactions across these rows are not only possible, but likely.
While Warschkow \textit{et al.} explored some of these reactions (corresponding to Fig.~\ref{fig:rxns1}b),\cite{warschkow2016reaction} we present a comprehensive report of these reactions and their reaction barriers in Fig.~\ref{fig:rxns1} and Fig.~\ref{fig:rxns2}, which we apply in the kinetic model used throughout this work.
We investigate PH$_3$ and PH$_2$ fragments either shedding hydrogen to the neighboring dimer row, or fully migrating on its own. 
We also simultaneously examine the same reaction when the neighboring dimer row is partially occupied, which leads to a uniform decrease of the reaction barrier. 

\begin{figure*}
\includegraphics[width=0.9\textwidth]{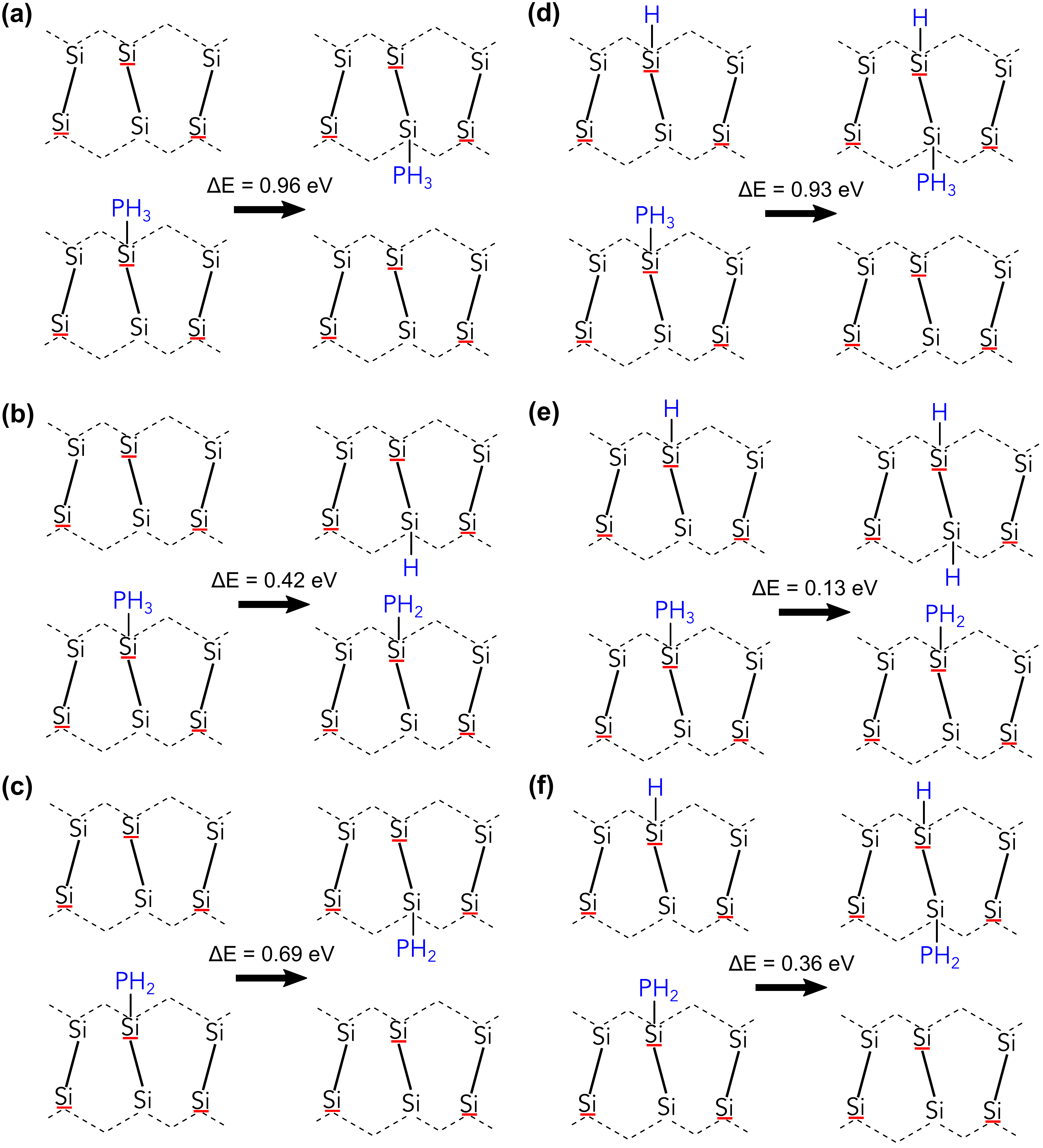}
\caption{ A summary of reaction barriers for PH$_x$ fragments either shedding hydrogen or migrating across dimer rows. These new reactions are included in the kinetic model used throughout the work. \label{fig:rxns1}}
\end{figure*}

Most notably, the barrier for the reaction shown in Fig.~\ref{fig:rxns1}b shows that a freshly adsorbed PH$_3$ losing its hydrogen across dimer rows has nearly the same barrier (0.42 eV) as losing within the same dimer row (0.46 eV).
This will lead to a significant spread of hydrogen across dimer rows as opposed to previous models where only reactions along the same dimer row are allowed.
We also find that the barrier for a PH$_2$ fragment jumping ship and moving to the neighboring silicon dimer row (0.69 eV) is only slightly higher than migration along the same dimer row (0.58 eV as predicted by Warschkow \textit{et al.}), making migration of the phosphorus atom itself to a different dimer row a small but distinct possibility.
In general, our calculations show that reactions that take place across multiple silicon dimer rows cannot be neglected in realistic models of multi-row patches. 

\begin{figure*}
\includegraphics[width=0.9\textwidth]{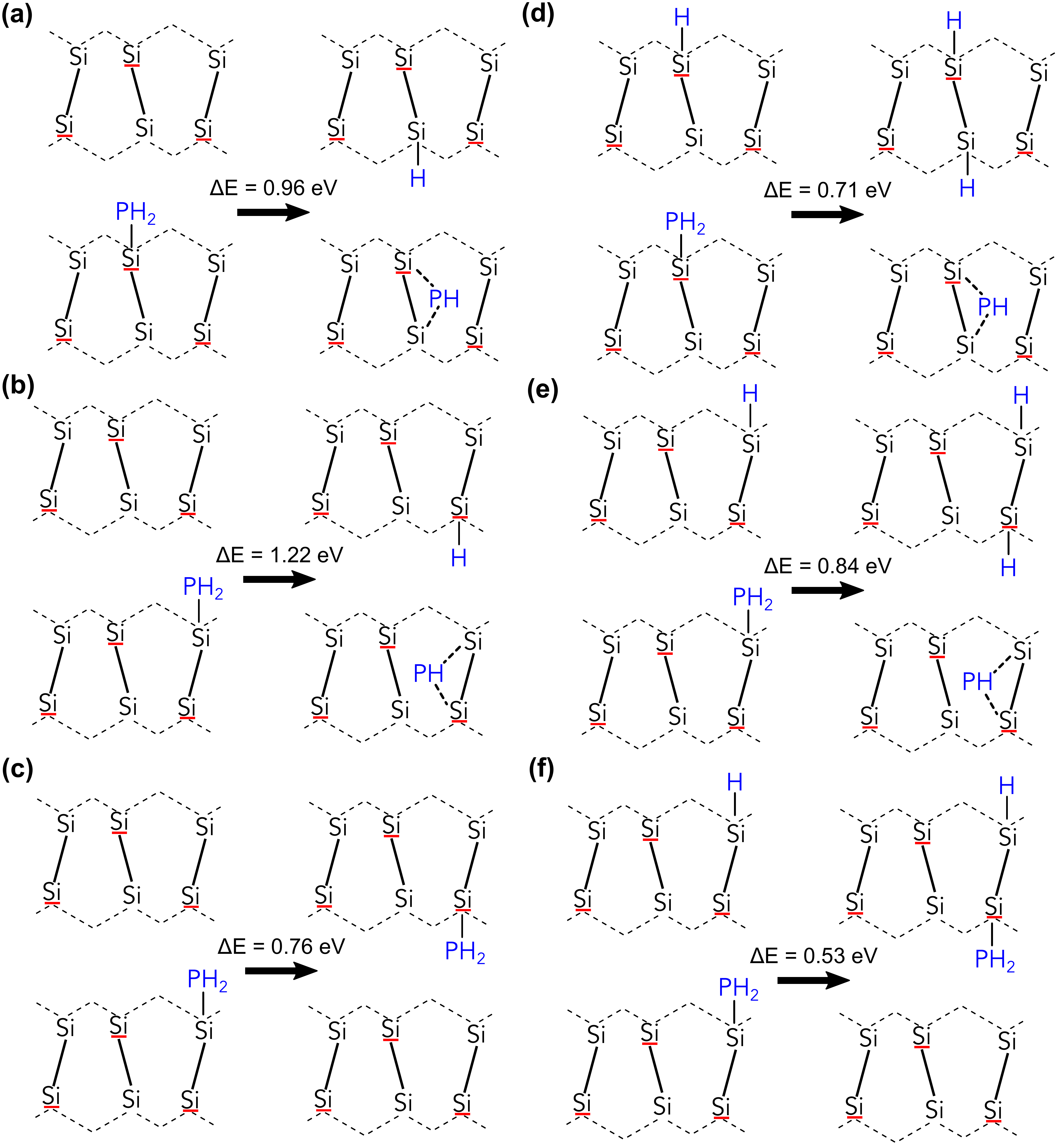}
\caption{ Continued summary of reaction barriers for PH$_x$ fragments either shedding hydrogen or migrating across dimer rows. These new reactions are included in the kinetic model used throughout the work.\label{fig:rxns2}}
\end{figure*}

\section*{Appendix B: Kinetic Monte Carlo Details}
\setcounter{equation}{0}
\setcounter{figure}{0}
\renewcommand{\theequation}{B\arabic{equation}}
\renewcommand{\thefigure}{B\arabic{figure}}
\label{sec:multi-row-rxns}

We use a Kinetic Monte Carlo model \cite{Bortz1975KMC,Gillespie1976KMC} as implemented in the \textsc{KMClib} package \cite{Leetmaa2014KMClib} to determine the probability of incorporation.
Our KMC model uses transition rates based on the Arrhenius equation $ \Gamma = A \exp{\Delta/k_{\rm B}T}$~\cite{arrhenius1889reaktionsgeschwindigkeit}, where $\Gamma$ is transition rate, $A$ is the attempt frequency, $\Delta$ is the reaction barrier found from our earlier DFT calculations, $k_{\rm B}$ is the Boltzmann constant, and $T$ is the temperature. 
We set all attempt frequencies $A$ to $10^{12}$ s$^{-1}$ as a reasonable order of magnitude estimate based on an analysis of attempt frequencies for the dissociation of phosphine on silicon~\cite{warschkow2016reaction}. 
We calculate the effusive flow rate of molecules landing on any particular silicon dimer as $\Phi_{effusion} =PA/\sqrt{2\pi m k_{\rm B}T}$, where $P$ is the pressure of the incoming precursor gas, $A$ is the area of impingement, taken here as a single silicon dimer, $m$ is the mass of the precursor gas.

Each KMC calculation is repeated 200 times with different random seeds, and the sample mean of the results is reported. 
We calculate error bars by assuming a binomial distribution of measured counts and using the standard error based on sample size.

\section*{Appendix C: 3 $\times$ 3 nm heatmaps of incorporation}
\setcounter{equation}{0}
\setcounter{figure}{0}
\renewcommand{\theequation}{C\arabic{equation}}
\renewcommand{\thefigure}{C\arabic{figure}}
\label{sec:multi-row-rxns}

\begin{figure*}
 \includegraphics[width=\columnwidth]{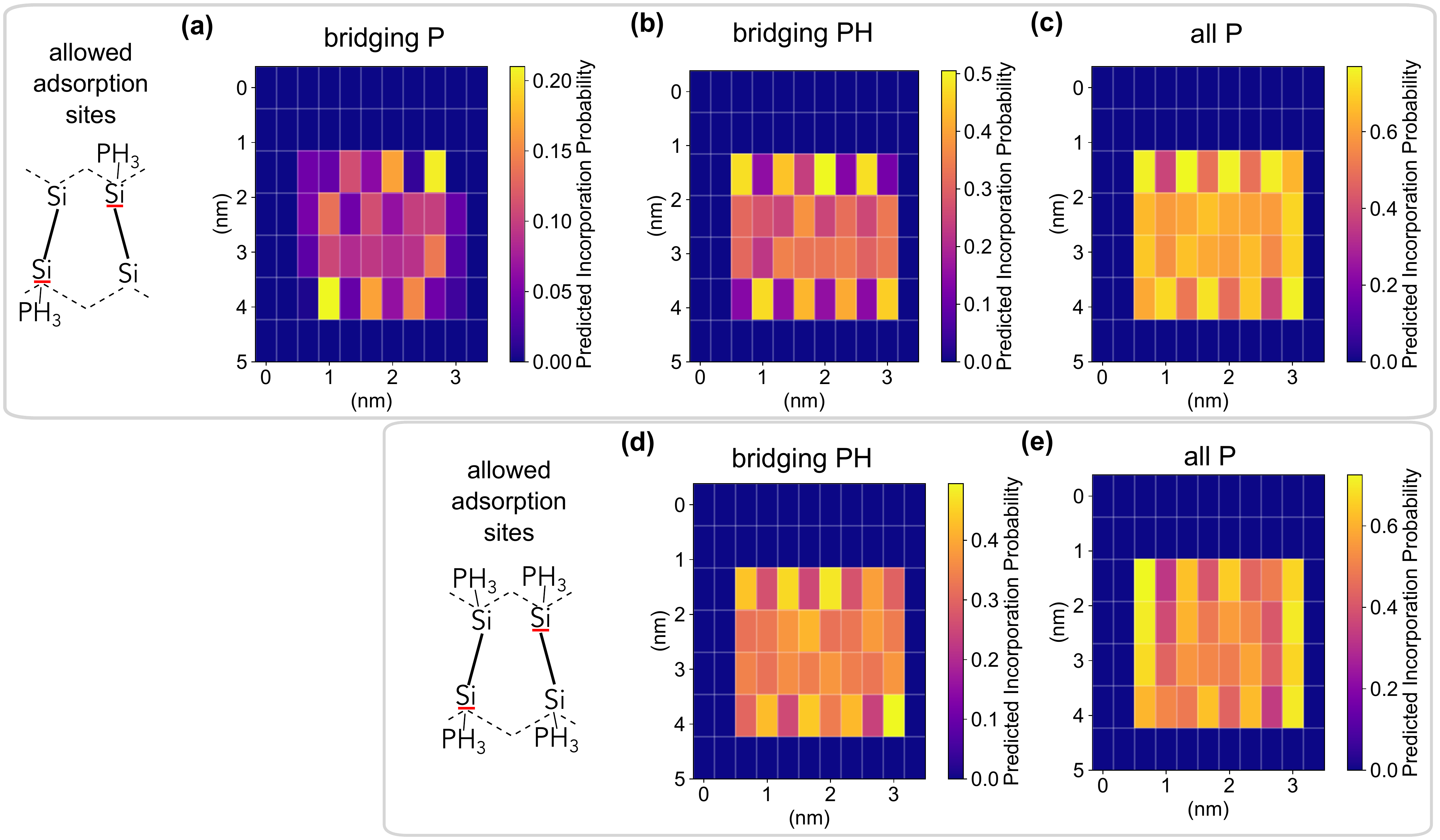}
\caption{Heatmaps of the incorporation of a 3 nm $\times$ 3 nm depassivated patch using a coverage of 3 L. 
For each heatmap, a different proxy for final incorporation is used. 
In (a) bridging P, (b) bridging PH, and (c) all P on the surface, we use simulations where phosphine can only adsorb on the lower end of dimers. 
In (d) bridging PH and (e) all P on the surface, we use simulations where phosphine molecules can adsorb on all depasivated silicon sites. 
The effect of ordering is steadily reduced as the proxy becomes less constricting and the adsorption sites more expansive.
}
\label{fig:3x3heatmap} 
\end{figure*}

We next expand our kinetic model to investigate larger windows to examine the spatial distribution of incorporation for wider $\delta$-doped regions.
In Fig.~\ref{fig:3x3heatmap}, we predict the spatial distribution of a 3 nm $\times$ 3 nm wide region which has been $\delta$-doped.
As noted in Sec.~\ref{sec:proxy-choice}, the choice within our model of which molecule to use as a proxy for eventual incorporation can lead to significant differences in both the level and position of incorporation events.

In Fig~\ref{fig:3x3heatmap}a, we use a bridging PH molecule as our proxy for incorporation. 
The most striking feature of the incorporation map here is the difference between the top and bottom dimer rows and the dimer rows in the middle of the structure.
The top and bottom dimer rows both exhibit a checkerboard pattern, with every other silicon dimer having significant levels of incorporation, while the other sites see dramatically lowered levels of incorporation.
This pattern can ultimately be attributed to the combination of the cross-dimer reaction illustrated in Fig.~\ref{fig:rxns1}b, and the alternating tilt of silicon dimers.
In the reconstructed silicon (100)-2$\times$1 surface, each atom of a silicon dimer either tilts up or down and the silicon dimer next to it will have the opposite arrangement of tilts.
Furthermore it is more favorable for a phosphine atom to adsorb on the lower end of the dimer, and within our model, we only allow this adsorption process, and not one at the higher end. 
This becomes important due to the low reaction barrier of a hydrogen dissociating from a PH$_3$ on one dimer row to land on the silicon atom of the nearby separate dimer row. 
In our model, this is the lowest reaction barrier for a PH$_3$ $\rightarrow$ PH$_2$ + H reaction, and thus highly favored. 
On the top and bottom rows, however, this means that it is only easy for half of the PH$_3$ molecules to dissociate: the half which have an open silicon dimer row next to it and are not site blocked by the hydrogen termination of the pattern. 
These PH$_2$ fragments go on to dissociate with high favorability and the shedded hydrogen further invade the nearby silicon dimers and block many dissociation pathway for these adsorbed PH$_3$ molecules. 
In the middle of the depassivated window, the phosphine can dissociate to neighboring dimer rows on either side, and thus the checkerboard effect of the alternating up-down tilt on incorporation is not observed.

While this pattern produces a striking predictions, it should be noted that while it is more favorable for a phosphine atom to adsorb on the lower end of a silicon dimer, it is still favorable on the upper end and likely to occur. 
Within our model, we have assumed that phosphine molecules will always adsorb at this lower end, but in reality this preference is less strong.
This, plus the tendency of dimer tilts to rapidly flip on the silicon surface, would likely lead to this checkerboard trend being, at best, dimly observed in real world devices. 
We verify this by also simulating systems where phosphine is allowed to adsorb on all silicon atoms with results shown in Fig.~\ref{fig:3x3heatmap}d and e. 
We indeed note that the checkerboard pattern is less well pronounced, although the predicted incorporation levels do not alter significantly.

\end{document}